\documentclass{aa}
\usepackage{graphicx}
\usepackage{times}
\begin{document}
\thesaurus{05                      
          (10.15.2 NGC\,581 (M\,103); 
           05.01.1;                 
           08.11.1;                 
           08.08.1;                 
           08.12.3)                
           }

\title{Photometric and Kinematic Studies of Open Star Clusters I:\\
 NGC\,581 (M\,103)}

\author{J\"org Sanner\inst{1}, Michael Geffert\inst{1}, Jens
  Brunzendorf\inst{2}, J\"urgen Schmoll\inst{3}}

\institute{Sternwarte der Universit\"at Bonn, Auf dem H\"ugel 71, 
           D--53121 Bonn, F.R. Germany \and
           Th\"uringer Landessternwarte Tautenburg, Sternwarte 5,
           D--07778 Tautenburg, F.R. Germany \and
           Astrophysikalisches Institut Potsdam, An der Sternwarte 16,
           D--14482 Potsdam, F.R. Germany}

\offprints{J\"org Sanner, jsanner@astro.uni-bonn.de}

\date{18 March 1999, 12 July 1999}

\maketitle

\markboth{J. Sanner et al.: NGC\,581 (M\,103)}{}

\begin{abstract}

We present CCD photometry and a proper motion study of the young open star
cluster, NGC\,581 (M\,103). Fitting isochrones to the colour magnitude
diagram, we found an age of $16 \pm 4$ Myr and a distance of roughly 3 kpc for
this cluster. The proper motion study identifies 77 stars of $V=14.5 \mbox{
  mag}$ or brighter to be cluster members. We combine membership determination
by proper motions and statistical field star subtraction to derive the IMF of
the cluster and find a quite steep slope of $\Gamma=-1.80$.


\end{abstract}

\section{Introduction}

\begin{figure}
\centerline{
\includegraphics[width=\hsize]{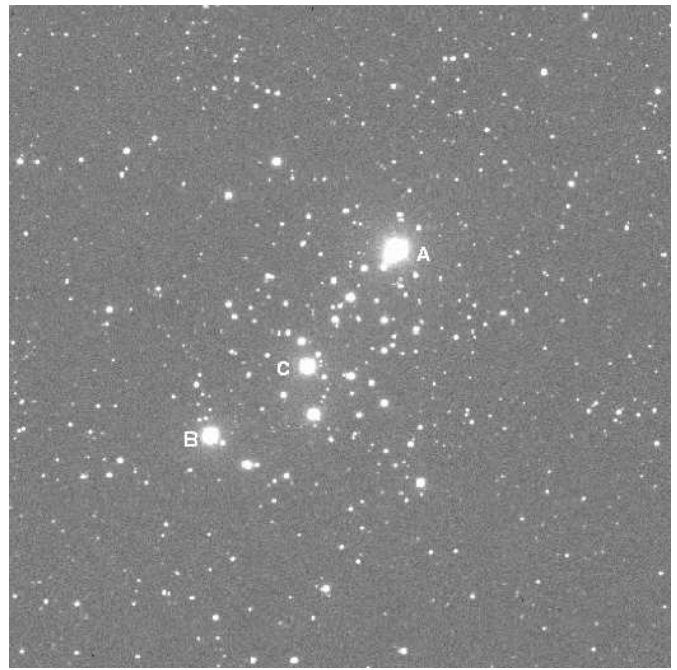}
}
\caption[]{\label{n0581bild} 60~s $V$ filter exposure of NGC\,581 (M\,103) taken with the 1m
  Cassegrain telescope of Hoher List observatory. The field of view reproduced
  here is approx. $15\arcmin \times 15\arcmin$. North is up and east to the
  left. The stars marked as A, B, and C are mentioned in
  Sect. \ref{n0581cmdanalysis}}
\end{figure}

The shape of the initial mass function (IMF) is an important parameter to
understand the fragmentation of molecular clouds and therefore the formation
and development of stellar systems. Besides studies of the Solar neighbourhood
(Salpeter \cite{salpeter}, Tsujimoto et al. \cite{tsuji}), work on star
clusters plays an important role (Scalo \cite{scalo1}), as age, metalicity,
and distance of all cluster stars can be assumed to be equal.

Most of the previous studies indicate that the IMF of a star cluster has the
shape of power laws 
\begin{equation}
N(m) \sim m^\Gamma
\end{equation}
within different mass intervals. The following typical values of their
exponents as given in Scalo (\cite{scalo2}):
\begin{eqnarray}
\Gamma=-1.3 & \mbox{ for } & m > 10 M_\odot, \nonumber\\
\Gamma=-1.7 & \mbox{ for } & 1 M_\odot < m < 10 M_\odot, \mbox{ and}\\
\Gamma=-0.2 & \mbox{ for } & m < 1 M_\odot. \nonumber
\end{eqnarray}
Knowledge of membership is essential to derive the IMF especially of
open star clusters, where the contamination of the data with field stars is a
major problem. Two methods for membership determination are in use nowadays
and each of them has its advantages and disadvantages:

\begin{itemize}
\item The classical method is to separate between cluster and field stars by
  their proper motions: All cluster stars can be expected to move in the same
  way, whereas the field stars show more widely spread and differently centred
  proper motions (see e.g.~the recent work of Francic \cite{francic}). For
  each star a membership probability can be specified. To obtain a sufficient
  epoch difference, old as well as recent photographic plates are needed to
  measure proper motions, so that this method is limited by the comparably
  poor sensitivity of the old plates.

\item With the introduction of CCD imaging to astronomy, statistical field
  star subtraction became more popular. Assuming (almost) identical field star
  distributions in the cluster region itself and the surrounding area,
  the distribution of the field stars can be subtracted from the one of the
  (contaminated) cluster area. This makes sense for fainter stars but with the
  bright stars one deals with statistics of small numbers.
\end{itemize}

Our work combines these two methods of membership determination:
The proper motions are investigated for the bright stars of the cluster,
whereas the fainter stars are treated with statistical field star
subtraction. From the cleaned data we derive the luminosity and mass functions
of the cluster.

NGC\,581 (M\,103), which is located at $\alpha_{2000.0}=1^h 33.2^m$,
$\delta_{2000.0}=+60 ^{\circ} 42 \arcmin$, was chosen as a first test object
for our technique because the cluster {\it and} a sufficiently large field
star region can be covered within the field of view of the telescope used. A
$V$ image of NGC\,581 is shown in Fig. \ref{n0581bild}.

In Sect. \ref{n0581photometry} we present our CCD photometry, and in
Sect. \ref{n0581propmot} a proper motion study of NGC\,581 and another cluster
which is located on the photographic plates, Trumpler 1. The resulting
colour magnitude diagram (CMD) is discussed in Sect. \ref{n0581feldsub},
leading to the determination of the IMF of NGC\,581 in
Sect. \ref{n0581imf}.

\section{CCD photometry}

\label{n0581photometry}

The photometry is based on 22 CCD frames taken in Johnson $B$ and $V$
filters at the 1m Cassegrain telescope of Hoher List Observatory. The
telescope was equipped with a focal reducing system and a 2k $\times$ 2k CCD
camera called HoLiCam (Sanner et al. \cite{holicam}), which has a pixel size
of $15 \mu \mbox{m} \times 15 \mu \mbox{m}$ and a resolution of $0.8 \arcsec
\mbox{pix}^{-1}$. The field of view covered in this configuration is a
circular area with a diameter of $28 \arcmin$. Information about the images
used for the photometry are summed up in Tab. \ref{n0581ccd}.

The images of equal exposure times were averaged, resulting in integrated
exposure times of 35 min in $V$ and 60 min in $B$ for the longest
exposures. The shorter images were used to gain information about the bright
stars which were saturated after longer exposure times. After standard image
processing the photometry was performed with DAOPHOT II (Stetson
\cite{stetson}) running under IRAF. After an error selection process, the data
were calibrated from instrumental to Johnson magnitudes using the
photoelectric sequence of Hoag et al. (\cite{navy}). Their standard stars as
well as our instrumental magnitude values and their deviations are given in
Table \ref{n0581wien}.

\begin{table}
\caption[]{\label{n0581ccd} Summary of the CCD images from the 1m Cassegrain
  telescope at Hoher List Observatory used for the photometry and
  the proper motion study}
\begin{tabular}{crcc}
\hline
filter & $t_{exp}$ & \multicolumn{2}{c}{number of exposures} \\
       & \multicolumn {1}{c}{[s]} & photometry & proper motions \\
\hline
V & 10 & 3 & 0 \\
V & 60 & 1 & 0 \\
V & 300 & 7 & 7 \\
B & 10 & 3 & 0 \\
B & 60 & 2 & 2 \\
B & 600 & 6 & 6 \\
\hline
\end{tabular}
\end{table}

\begin{table}
\label{n0581wien}
\caption[]{Internal standard stars of Hoag et al. (\cite{navy}) with the
  deviations of the computed from the catalogue magnitudes. Stars 1 and 2 were
  saturated even in the shortest exposures}
\begin{tabular}{rrrrr}
\hline
\multicolumn{1}{c}{no.} & \multicolumn{1}{c}{$V$} & \multicolumn{1}{c}{$B-V$} & \multicolumn{1}{c}{$\Delta V$} & \multicolumn{1}{c}{$\Delta (B-V)$} \\
 & \multicolumn{1}{c}{[mag]} & \multicolumn{1}{c}{[mag]} &
 \multicolumn{1}{c}{[mag]} & \multicolumn{1}{c}{[mag]}\\
\hline
 3 &  9.09 & +0.21 & $-0.020$ & $+0.039$ \\
 4 & 10.45 & +0.24 & $-0.038$ & $+0.007$ \\
 5 & 10.59 & +0.17 & $-0.013$ & $+0.002$ \\
 6 & 10.81 & +1.93 & $+0.004$ & $-0.034$ \\
 7 & 11.22 & +0.23 & $-0.028$ & $+0.015$ \\
 8 & 11.35 & +0.26 & $-0.002$ & $+0.013$ \\
 9 & 11.76 & +0.22 & $-0.016$ & $-0.022$ \\
10 & 11.84 & +0.25 & $-0.031$ & $+0.021$ \\
11 & 12.34 & +0.10 & $+0.083$ & $-0.105$ \\
12 & 12.76 & +0.24 & $+0.090$ & $-0.047$ \\
13 & 13.17 & +0.30 & $-0.016$ & $+0.023$ \\
14 & 13.21 & +0.26 & $+0.012$ & $-0.119$ \\
15 & 13.27 & +0.44 & $+0.034$ & $+0.019$ \\
16 & 13.46 & +0.30 & $-0.009$ & $+0.031$ \\
17 & 13.47 & +0.35 & $-0.028$ & $+0.075$ \\
18 & 13.59 & +0.77 & $-0.001$ & $+0.060$ \\
19 & 13.76 & +0.30 & $-0.008$ & $-0.010$ \\
\hline
\multicolumn{3}{c}{standard deviations} & $0.036$ & $0.051$\\
\hline
\end{tabular}
\end{table}

We applied the following equations to transform instrumental to apparent
magnitudes:
\begin{eqnarray}
 v-V &=& a_0 - a_1 \cdot (B-V)\\
(b-v)-(B-V) &=&  a'_0 - a'_1 \cdot (B-V)
\end{eqnarray}
with
\begin{eqnarray}
a_0=5.536 \pm 0.04, & & a_1=0.094 \pm 0.02\\
a'_0=2.418 \pm 0.04, & & a'_1=0.139 \pm 0.02
\end{eqnarray}
where $B$ and $V$ represent apparent and $b$ and $v$ instrumental
magnitudes, respectively. Mean photometric errors in different magnitude
intervals are given in Table \ref{n0581photerrors}. One can see that the
errors for $B$ increase more rapidly as a consequence of HoLiCam's poorer
sensitivity in blue wavelengths. Although the total exposure time in $B$ was
almost twice as large as in $V$, the limiting magnitude of the photometry is
defined by the $B$ images.

\begin{table}
\caption[]{\label{n0581photerrors}Photometric errors in different magnitude
  ranges}
\begin{tabular}{r@{$V$}lcc}
\hline
\multicolumn{2}{c}{range} & $\Delta V [\mbox{mag}]$ & $\Delta B[\mbox{mag}]$ \\
\hline
         & $< 12 \mbox{ mag}$ & 0.007 & 0.009 \\
 $12 \mbox{ mag} <$  & $< 16 \mbox{ mag}$ & 0.012 & 0.021 \\
 $16 \mbox{ mag} <$  &        & 0.028 & 0.092 \\
\hline
\end{tabular}
\end{table}

\begin{figure}
\centerline{
\includegraphics[width=\hsize]{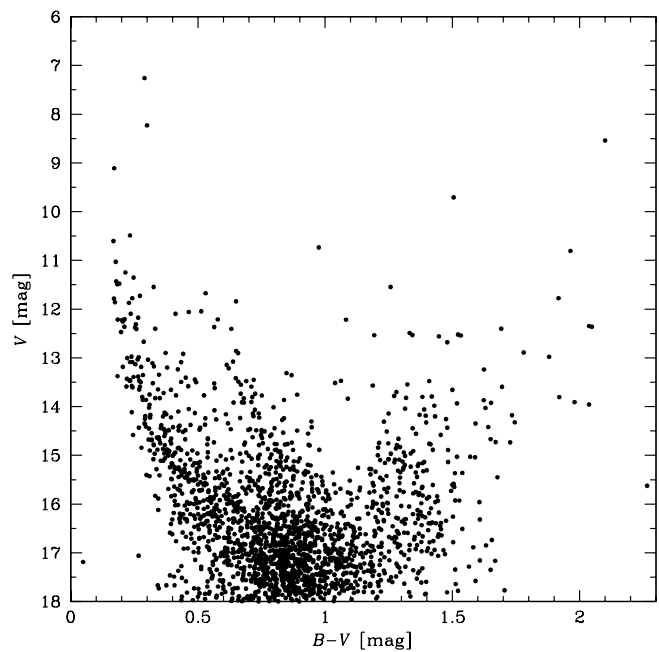}
}
\caption[]{\label{n0581cmd} Colour magnitude diagram of all objects detected
  on the CCD frames of NGC\,581. Note the two main sequence structures and the
  scattered red giant branch. The origin of these features is discussed in
  Sect. \ref{n0581feldsub}}
\end{figure}

From these data, we determined a CMD which is shown in
Fig. \ref{n0581cmd}. It represents a total of 2134 stars for which both $V$
and $B-V$ magnitudes are available. We present the photometric data of
all these objects in Table \ref{n0581cdsphot}.

\begin{table}
\caption[]{\label{n0581cdsphot} List of the photometric data of all stars
  measured in the CCD field of NGC\,581. Star numbers higher than 10,000
  account for objects for which proper motions  were determined, too. Only the
  ten brightest stars for which both photometry and proper motions are
  available are listed here, the complete table is available online at the CDS
  Strasbourg archive}
\begin{tabular}{rrrrr}
\hline
\multicolumn{1}{c}{No.} & \multicolumn{1}{c}{$x$} & \multicolumn{1}{c}{$y$} &
\multicolumn{1}{c}{$V$} & \multicolumn{1}{c}{$B-V$} \\
&  &  & \multicolumn{1}{c}{[mag]} & \multicolumn{1}{c}{[mag]} \\
\hline
14380 & 1024.178 & 1122.116 &   10.603  &   0.168\\
14561 & 1136.343 & 1327.688 &   10.806  &   1.964\\
14503 & 1099.685 & 1056.475 &   11.028  &   0.177\\
14411 & 1046.961 & 1088.074 &   11.432  &   0.179\\
14288 &  972.818 & 1087.454 &   11.478  &   0.191\\
13899 &  697.822 &  446.703 &   11.674  &   0.530\\
13894 &  708.770 & 1589.465 &   11.776  &   1.917\\
14223 &  925.276 & 1073.279 &   11.776  &   0.242\\
14341 & 1003.143 & 1165.883 &   11.785  &   0.170\\
14928 & 1391.922 & 1103.418 &   11.871  &   0.229\\
\hline
\end{tabular}
\end{table}

The CMD shows two main sequence features and a scattered giant branch in a
colour range around $B-V \simeq 1.4 \mbox{ mag}$. More detailed analysis of
the CMD is being presented in Sect. \ref{n0581feldsub}.

\section{Proper motion study}

\label{n0581propmot}

\subsection{Data reduction}

For the proper motion study eight photographic plates from the Bonn
Doppelrefraktor (until 1965 located in Bonn, thereafter at Hoher List
Observatory) were used, covering an epoch difference of 81 years. The 16 cm
$\times$ 16 cm plates of the $D=0.3 \mbox{ m}$, $f=5.1 \mbox{ m}$ instrument
represent a region of $1.6 ^{\circ} \times 1.6 ^{\circ}$. The plates were
digitized with the PDS machines of the Astronomi\-sches Institut M\"unster and
the Tautenburg Plate Scanner (TPS) of Th\"uringer Landessternwarte Tautenburg
(Brunzendorf \& Meu\-sing\-er \cite{TPS}). The positions gained with DAOPHOT II
from 15 CCD frames were added to the plate data. Tables \ref{n0581ccd} and
\ref{n0581platten} give an overview of the material included in the proper
motion study.

\begin{table}
\caption[]{\label{n0581platten} Photographic plates from the Bonn
  Doppelrefraktor used for the proper motion study. The plates marked with MS
  were digitized with the PDS machines in M\"unster, the ones with T with TPS
  in Tautenburg. For comparison, both digitizations of R 0285 (M\"unster and
  Tautenburg) were used during data reduction}
\begin{tabular}{llrr}
\hline
\multicolumn{1}{c}{plate no.} & \multicolumn{1}{c}{date} & \multicolumn{1}{c}{$t_{exp}$} & \multicolumn{1}{c}{scan}\\
 & & \multicolumn{1}{c}{$[\mbox{min}]$} & \\
\hline
R 0285 &  23.01.1917 & 30 & MS/T\\
R 0294 &  09.02.1917 & 30 & MS\\
R 0295 &  09.02.1917 &  3 & T\\
R 1281 &  03.12.1977 & 60 & MS\\
R 1291 &  05.12.1977 & 64 & MS\\
R 1295 &  19.12.1977 &  4 & T\\
R 1298 &  19.12.1977 &  4 & T\\
R 1899 &  24.03.1998 & 60 & T\\
\hline
\end{tabular}
\end{table}

The celestial positions of the stars were determined from the plate
coordinates with respect to six HIPPARCOS stars (ESA \cite{hipp}) using an
astrometric software package developed by Geffert et al. (\cite{geffert}). We
obtained good results using quadratic polynomials (6 plate constants) for
transforming $(x,y)$ to $(\alpha,\delta)$ for the photographic plates and
cubic polynomials (10 constants) for the CCD images, respectively. The mean
positional deviations of the HIPPARCOS stars after the first reduction step
were of the order of $0.1 \arcsec$ in both right ascension and
declination. Using the output positions and proper motions of each step as the
basis of the next run, we derived a stable solution of proper motions of a
total of 2,387 stars on the whole field after four iterations with a mean
error for the proper motions of approx. $1.1 \mbox{ mas\,yr}^{-1}$ in both
coordinates. The differences between our measurements and the HIPPARCOS data
(``observed -- calculated'' or $O-C$ values) are listed in Table
\ref{n0581hipp}. Compared with our measurements, HIPPARCOS star no. 7155
showed high deviations --- probably caused by a double star nature of this
object (Wielen et al. \cite{wielen}) --- and was therefore excluded before the
data reduction. We present the proper motions of all stars in Table
\ref{n0581cdseb}.

\begin{table}
\caption[]{\label{n0581hipp}''$O-C$'' values of the measurements of the
  HIPPARCOS stars which were used for the transformation from pixel to
  celestial coordinates}
\begin{tabular}{rrrrr}
\hline
\multicolumn{1}{c}{star no.} & \multicolumn{1}{c}{$\Delta \alpha$} & \multicolumn{1}{c}{$\Delta \delta$} & \multicolumn{1}{c}{$\Delta \mu_\alpha \cos \delta$}
& \multicolumn{1}{c}{$\Delta \mu_\delta$}\\
 & \multicolumn{1}{c}{$[\arcsec]$} & \multicolumn{1}{c}{$[\arcsec]$} &
 \multicolumn{1}{c}{[mas\,yr$^{-1}$]} & \multicolumn{1}{c}{[mas\,yr$^{-1}$]} \\
\hline
6979 & $ 0.027$ & $-0.020$ & $ 0.22$ & $-0.95$ \\
7640 & $-0.006$ & $-0.080$ & $-2.20$ & $-1.38$ \\
7497 & $-0.006$ & $-0.130$ & $-0.91$ & $-1.72$ \\
6793 & $-0.003$ & $-0.020$ & $ 1.17$ & $ 2.40$ \\
6927 & $-0.011$ & $-0.080$ & $ 1.93$ & $-1.10$ \\
6841 & $ 0.050$ & $-0.330$ & $-0.85$ & $ 0.50$ \\
\hline
\end{tabular}
\end{table}

\begin{table}
\caption[]{\label{n0581cdseb} List of all proper motions determined from
    the photographic plates of NGC\,581. The positions are given for the epoch 
    1950.0 in the equinox 2000.0 coordinate stystem. The stellar numbers are
    the same as in Table \ref{n0581cdsphot}. Only the proper motions of the
    same stars as in Table \ref{n0581cdsphot} are presented here, the complete
    table is available online at the CDS Strasbourg archive}
\begin{tabular}{ccccc}
\hline
No. & $\alpha$ & $\delta$ & $\mu_\alpha \cos \delta$ & $\mu_\delta$ \\
 & [h m s] & [$^\circ \mbox{ } \arcmin \mbox{ } \arcsec$ ] & [mas yr$^{-1}$] & [mas yr$^{-1}$] \\
\hline
14380 & 1 33 21.800 & +60 40 12.25 & $-0.29$ & $+0.74$ \\
14561 & 1 33 33.908 & +60 42 59.62 & $-1.00$ & $-1.74$ \\
14503 & 1 33 30.184 & +60 39 19.62 & $-0.68$ & $-0.53$ \\
14411 & 1 33 24.347 & +60 39 44.88 & $+0.10$ & $+2.95$ \\
14288 & 1 33 16.193 & +60 39 43.50 & $-0.51$ & $+1.27$ \\
13899 & 1 32 46.876 & +60 31 02.94 & $+9.07$ & $+6.72$ \\
13894 & 1 32 46.417 & +60 46 26.62 & $+0.04$ & $-1.44$ \\
14223 & 1 33 10.977 & +60 39 31.50 & $+0.16$ & $-0.18$ \\
14341 & 1 33 19.427 & +60 40 47.50 & $-0.25$ & $-1.30$ \\
14928 & 1 34 02.261 & +60 40 00.25 & $-0.87$ & $-0.55$ \\
\hline
\end{tabular}
\end{table}

\subsection{NGC\,581}

\label{n0581pm}

\begin{figure}
\centerline{
\includegraphics[width=\hsize]{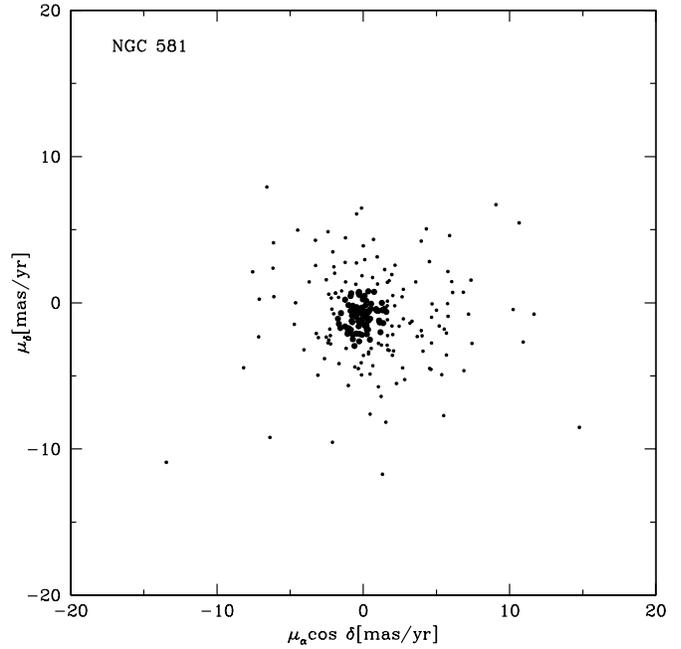}
}
\caption[]{\label{n0581vppd} Vector point plot diagram of the stars in the region
  of NGC\,581. Note that it is difficult to distinguish between the cluster
  proper motions and the centre of the field stars' proper motions. The stars
  with a membership probability of less than 0.8 are indicated by small, the
  others by larger dots. The width of the distribution of the stars with a
  high membership probability is of the order of $1 \mbox{ mas\,yr}^{-1}$
  which coincides with the standard deviation of the proper motion of a single
  star}
\end{figure}

228 stars were located within $10 \arcmin$ from the centre of NGC\,581. Only
these shown in Fig. \ref{n0581vppd} were taken into account for the vector
point plot diagram. Due to the cluster's large distance (see
Sect. \ref{n0581cmdanalysis}), the centres of the distributions of field and
cluster stars are more or less the same. Therefore, the separation between
field and cluster stars is not very apparent.

\begin{figure}
\centerline{
\includegraphics[width=\hsize]{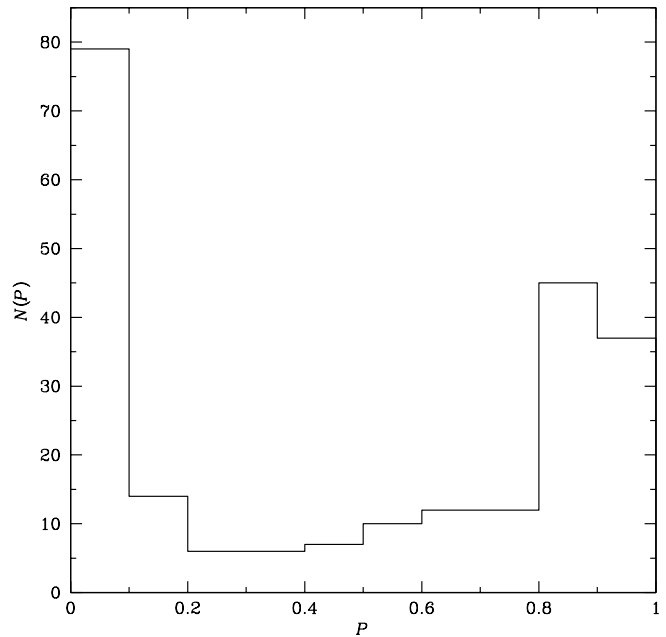}
}
\caption[]{\label{n0581phist} Histogram of the membership probability for the
  stars from Fig. \ref{n0581vppd}}
\end{figure}

\begin{figure}
\centerline{
\includegraphics[width=\hsize]{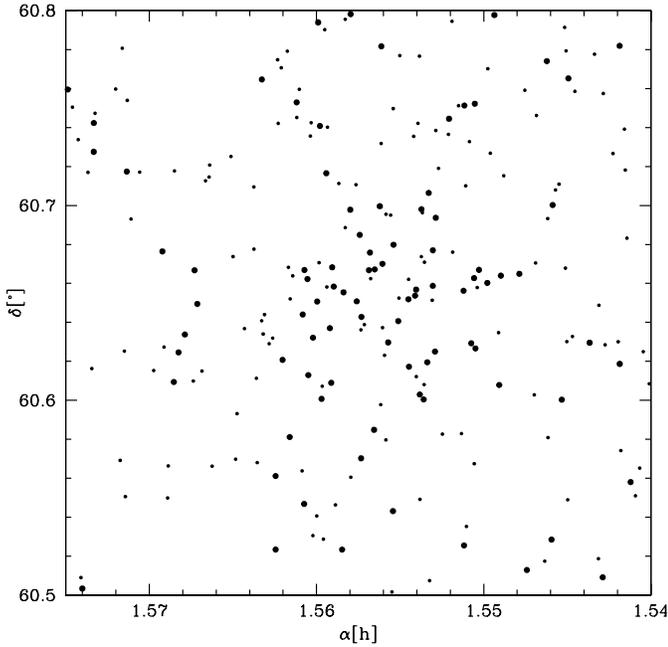}
}
\caption[]{\label{n0581pos} Diagram of the $(\alpha,\delta)$ positions of the
  cluster member and non-member stars due to the proper motions. The
  non-member stars are indicated by small, member stars by larger dots. Note
  that the field stars are distributed homogeneously over the whole
  area, however, the member stars are located in a larger than expected
  area. This point is discussed in Sect. \ref{n0581pm}}
\end{figure}

According to the method presented by Sanders (\cite{sanders}), we fitted a
sharp (for the members) and a wider spread (for the field stars) Gaussian
distribution to the distribution of the stars in the vector point plot
diagram. We computed the parameters of the two distributions with a maximum
likelihood method. From the values of the distribution at the location of the
stars in the diagram we derived the membership probabilities. Due to the small
difference between the centres of the two distributions the stars with proper
motions far away from the maximum are clearly identified as non-members,
whereas it is difficult to decide which objects of the central region do
belong to the cluster and which not. This is represented in the histogram for
the membership probabilities (Fig.\ref{n0581phist}) with a clear peak at $P
\simeq 0$ and a much less pronounced increase towards
$P=1$. Fig. \ref{n0581pos} shows the positions of the member and non-member
stars.

With this method 77 stars were classified to be members of NGC\,581 with a
probability of at least 0.8, while 151 objects show a lower membership
probability. The average proper motions for cluster stars are:
\begin{eqnarray*}
\mu_\alpha \cos \delta &=& (0.11 \pm 1.07) \mbox{ mas\,yr}^{-1}\\
\mu_\delta &=& (-0.95 \pm 1.24) \mbox{ mas\,yr}^{-1}
\end{eqnarray*}
and for the field stars:
\begin{eqnarray*}
\mu_\alpha \cos \delta &=& (1.14 \pm 4.46) \mbox{ mas\,yr}^{-1}\\
\mu_\delta &=& (-0.91 \pm 3.77) \mbox{ mas\,yr}^{-1}.
\end{eqnarray*}
Although the values are very similar, the standard deviations show a large
difference between field and cluster stars. We should remark that the "true"
members might be more concentrated towards the cluster centre, but because of
the problems in dividing between cluster and field stars, some non-members
might be taken for member stars and vice versa.

\subsection{Trumpler\,1}

Near the edge of the photographic plates we found the open star cluster
Trumpler\,1 for which we attempted a proper motion study, too, as a by-product
of our work.

\begin{figure}
\centerline{
\includegraphics[width=\hsize]{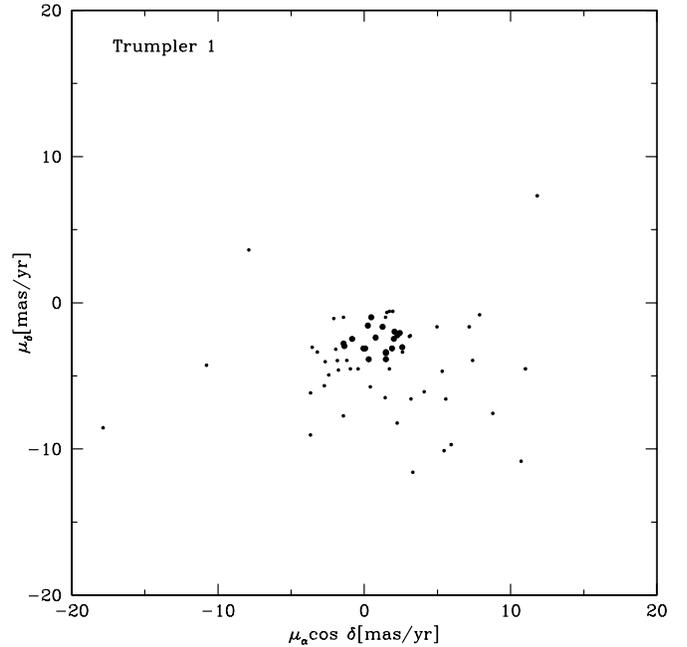}
}
\caption[]{\label{tr1vppd} Vector point plot diagram of the stars in the region
  of the open star cluster Trumpler\,1. Again, the stars with a membership
  probability of less than 0.8 are indicated by small, the others by large
  dots. Be aware that this proper motion study is of less reliability than
  the previous one of NGC\,581, as it is based only on photographic plates, and
  the cluster is located close to the edge of the plates}
\end{figure}

In a region of $0.15^{\circ}$ around the centre of the cluster at
$\alpha_{2000}=01^{\mbox{h}}35^{\mbox{m}}.7$, $\delta_{2000}=+61^{\circ} 17
\arcmin$, we detected only 64 stars on the base of the photographic
plates. CCD photometry is not available. A membership analysis with the same
methods as described above resulted in average proper motions of
\begin{eqnarray*}
\mu_\alpha \cos \delta &=& (0.59 \pm 1.71) \mbox{ mas\,yr}^{-1}\\
\mu_\delta &=& (-2.68 \pm 1.19) \mbox{ mas\,yr}^{-1} 
\end{eqnarray*}
for the cluster members and
\begin{eqnarray*}
\mu_\alpha \cos \delta &=& (1.41 \pm 5.73) \mbox{ mas\,yr}^{-1}\\
\mu_\delta &=& (-4.59 \pm 3.93) \mbox{ mas\,yr}^{-1}.
\end{eqnarray*}
for the field stars. We present a vector point plot diagram in
Fig. \ref{tr1vppd}.

It should be mentioned that these proper motions are less accurate than
our results for NGC\,581, as Trumpler\,1 is located close to the edge of the
plates and we did not include CCD positions in our analysis. However, the
results indicate that Trumpler\,1 and NGC\,581 have not only roughly the same
ages and distances (see Phelps \& Janes \cite{phelps1}), but --- within the
errors --- the same absolute proper motions. Therefore, the two objects may be
taken as a candidate for a binary star cluster (see e.g. Subramaniam et
al. \cite{subram}) as they are known for the Magellanic Clouds (Dieball \&
Grebel \cite{dieball}).

\section{Analysis of the colour magnitude diagram}

\label{n0581feldsub}

\subsection{Field star subtraction}

The CMD derived in Sect. \ref{n0581photometry} is contaminated with field
stars. Before trying to analyse the CMD it is essential to subtract these
stars to emphasise the features which belong to NGC\,581 and to prepare the
data for the determination of the cluster IMF.

In the magnitude range covered by the proper motion study it is possible to
distinguish between field and cluster stars using the information of the
membership probabilities from Sect. \ref{n0581propmot}. The proper motion
study is complete down to $V=14 \mbox{ mag}$ so that until this point, a
membership probability of 0.8 or higher is considered a suitable criterion for
the definition of the members of NGC\,581.

Below this limit, the star numbers are high enough to justify a statistical
field star subtraction. For this, the CCD field was divided into two regions
of the same area of $1.3 \cdot 10^6$ pixels: a circular region with the star
cluster in its centre (radius: 654 pix. or $18 \arcmin$) and a ring encircling
this region (outer radius: 925 pix. or $26 \arcmin$). Under these assumptions,
all cluster members are surely located in the central area. For both parts
separate CMDs were computed. Each of them was divided into rectangular bins
with a length of $0.5 \mbox{ mag}$ in magnitude and $0.1 \mbox{ mag}$ in
colour. (Variation of the bin sizes did not affect the results.) The numbers
of stars in the field CMD cells were determined and as many stars of the
corresponding cell of the inner region CMD were randomly chosen and
removed. Assuming a homogeneous distribution of field stars over the whole
field of view, the resulting CMD, which is presented in
Fig. \ref{n0581cmdhaufen}, represents only the cluster members.

\begin{figure}
\centerline{
\includegraphics[width=\hsize]{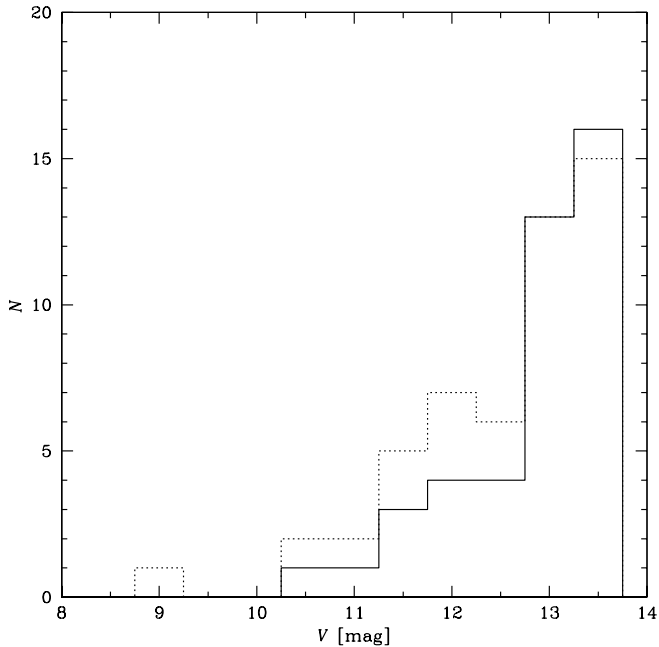}
}
\caption[]{\label{n0581vergleich} Luminosity functions down to $V=14 \mbox{
    mag}$ according to field star subtraction with proper motions (solid line)
  and statistical subtraction (dotted line). Statistical field star
  subtraction for the brighter stars would have lead to 9 additional cluster
  members and therefore to a shallower IMF}
\end{figure}

To illustrate the advantage of our method, Fig. \ref{n0581vergleich} shows the
luminosity functions obtained with the proper motions and the statistical
field star subtraction, respectively, in the range of the photographic plates
($V<14 \mbox{ mag}$). It can be seen that the proper motion analysis leads to
less member stars than the statistical subtraction.

\subsection{Morphology of the cluster CMD and age determination}

\label{n0581cmdanalysis}

The brightest stars in the field of NGC\,581 (marked with letters A to C in
Fig. \ref{n0581bild}) were saturated even in the shortest exposures so that no
direct photometric information was available for them. However, the three
stars were the subject of previous photoelectric studies (Hoag et
al. \cite{navy}, Purgathofer \cite{wien}) so that their $V$ magnitudes and
$B-V$ colours could be added to our CMD manually. Star A was too bright to be
measured on the photographic plates with a sufficient accuracy so that we
could not derive its proper motion, either. However, since star A is a
HIPPARCOS star (ESA \cite{hipp}, star no. 7232), we could get its proper
motion from this source. Calculating its membership probability led to a value
of 0.89. Star B has a proper motion far away from the one of the cluster so
that it can be considered a non-member, while star C belongs to NGC\,581 with
a probability of 0.90.

After field star subtraction, we found three red stars with $B-V \approx 2
\mbox{ mag}$ from $V \approx 10.8 \mbox{ mag}$ to $V \approx 12.4 \mbox{
  mag}$. Compared with the well populated main sequence in the same magnitude
range, it seems obvious that these objects cannot be cluster members. We
assume these objects to be field stars which are classified as member stars as
a consequence of the small difference between field and cluster proper
motions.

The secondary main sequence and giant branch like structures completely
vanished within the range of the proper motions. Below $V=14 \mbox{ mag}$, we
still find some stars remaining in the region of these features. Since the
vector point plot diagram does not give any evidence for a second cluster in
the same line of sight, we assume that these stars belong to the field star
population(s). It is very unlikely, too, that the remaining stars are pre main
sequence cluster members, because at the age of NGC\,581 (see the following
paragraph) this kind of objects should be much closer to the main sequence
than the stars in the CMD are (Iben \cite{iben}). Therefore we assume these
stars to be remnants of the field star subtraction due to the imperfect
statistics of the sample.

\begin{figure}
\centerline{
\includegraphics[width=\hsize]{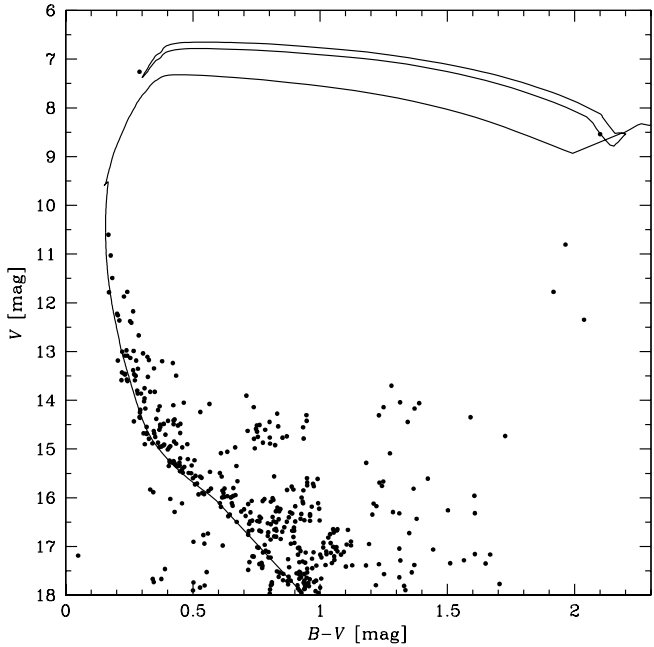}
}
\caption[]{\label{n0581cmdhaufen} Colour magnitude diagram of all members of
  NGC\,581 as determined with the proper motions ($V<14 \mbox{ mag}$) and the
  statistical field star subtraction ($V>14 \mbox{ mag}$). The parameters of
  the isochrone plotted in the diagram are listed in Table
  \ref{n0581params}. Note that the brightest star of Fig. \ref{n0581cmd} is
  not a member of the cluster}
\end{figure}

We fitted isochrones of the Geneva group (Schaller et al. \cite{schaller}) to
the resulting CMD from which we derived the distance modulus, reddening, age,
and metalicity of NGC\,581. The best fitting isochrone is plotted into the
cleaned CMD of Fig. \ref{n0581cmdhaufen}. The parameters of the selected
isochrone are shown in Table \ref{n0581params}. Isochrones of the Padua group
(Bertelli et al. \cite{padua}) were fitted to the CMD for comparison and led
to the same set of parameters.

\begin{table}
\caption{\label{n0581params} Parameters of NGC\,581 as derived from isochrone
  fitting to the colour magnitude diagram. The errors of distance modulus
  and reddening are estimated, the uncertainty of the age is derived from the
  comparison with the isochrones neighbouring the selected one}
\begin{tabular}{lr@{ = }l}
\hline
distance modulus & $(m-M)_0$     & $12.3 \pm 0.1 \mbox{ mag}$\\
i.e. distance    & $r$       & $2884 \pm 130 $ pc  \\
reddening        & $E_{B-V}$ & $0.41 \pm 0.02 \mbox{ mag}$\\
age              & $\log t$  & $7.2 \pm 0.1$\\
i.e.             & $t$       & $16 \pm 4 \mbox{ Myr}$\\
metalicity       & $Z$       & $0.02$\\
\hline
\end{tabular}
\end{table}

\subsection{Completeness correction}

\label{n0581compl}

To obtain comprehensive luminosity and initial mass functions, the data have
to be corrected for completeness. As crowding is not a problem in our images,
the completeness in the field and cluster regions are the same so that it is
not necessary to correct for completeness before the field star subtraction.

\begin{figure}
\centerline{
\includegraphics[width=\hsize]{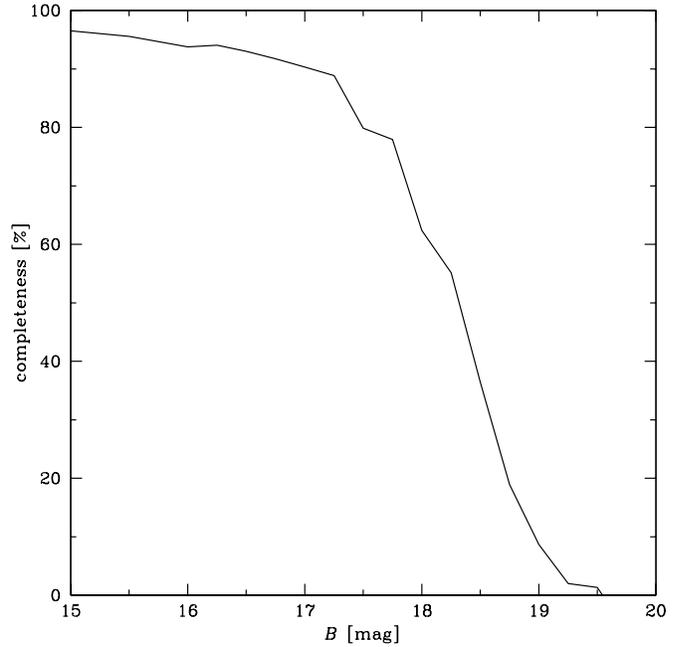}
}
\caption[]{\label{n0581complete} Diagram of the completeness of our data as a
  function of $B$ magnitudes. Down to $B=17.25 \mbox{ mag}$, the sample is
  more than $90 \%$ complete, down to one magnitude fainter, we detect a
  completeness of more than $50 \%$. After this point, the completeness
  sharply drops until no more stars are being detected at $B=19.5 \mbox{
  mag}$. For our further investigations we skip all stars fainter than $B=18
  \mbox{ mag}$ which corresponds to a completeness level of higher than $60
  \%$}
\end{figure}

With artificial star experiments using the DAOPHOT II routine {\tt addstar} we
derived the completeness function shown in Fig. \ref{n0581complete}. As the $V$
images reach down to much fainter magnitudes, these experiments were only
performed with the $B$ exposures. Down to $B=17.25 \mbox{ mag}$ the sample is
more than $80 \%$ complete with a sharp drop afterwards to almost $0 \%$ at
$B=19 \mbox{ mag}$. Figure \ref{n0581complete} shows that the completeness
level of $60 \%$ is reached around $B=18 \mbox{ mag}$, so that we assume this
value to be a reasonable cut-off for our studies. With a main sequence star
colour of $B-V=0.8 \mbox{ mag}$ for $B=18 \mbox{ mag}$, this corresponds to a
limiting magnitude of $V=17.2 \mbox{ mag}$.

\subsection{Initial mass function}

\label{n0581imf}

\begin{figure}
\centerline{
\includegraphics[width=\hsize]{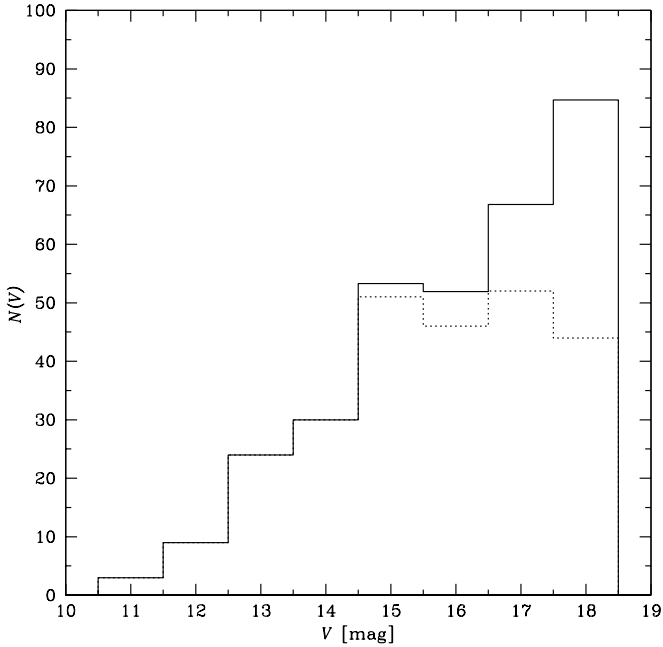}
}
\caption[]{\label{n0581lkf} Luminosity function of NGC\,581. The histogram
  obtained after completeness correction is printed with solid, the
  uncorrected one with dotted lines}
\end{figure}

With the stars of the cleaned CMD we determined the luminosity
(Fig. \ref{n0581lkf}) and initial mass functions of NGC\,581 after deleting
all objects outside a $\Delta (B-V)=0.3 \mbox{ mag}$ wide strip around the main
sequence and above the turnover of the isochrone around $V=10 \mbox{ mag}$
from the CMD and applying the completeness correction described in
Sect. \ref{n0581compl}. As mentioned in Sect. \ref{n0581pm}, some field stars
will have stayed in the sample while a few cluster stars might have been
rejected. However, we assume these two effects do not influence the IMF of
NGC\,581.

From the initial stellar masses given in the Geneva isochrone data, we
computed a mass-luminosity relation represented by a $6^{\mbox{th}}$ order
polynomial:
\begin{equation} \label{n0581mlr}
m[M_\odot]=\sum_{i=0}^6 d_i \cdot V^i[mag]
\end{equation}
with the following parameters:
\begin{eqnarray*}
d_0&=&-523.815\\
d_1&=&+221.123\\
d_2&=&-36.347\\
d_3&=&+3.055\\
d_4&=&-0.140\\
d_5&=&+0.003\\
d_6&=&-0.0000327.
\end{eqnarray*}
A $5^{\mbox{th}}$ order polynomial still did not fit the faint part of the
mass-luminosity relation well enough. With Eq. (\ref{n0581mlr}) we derived the
initial masses of the stars classified as objects of NGC\,581 on the base of
their $V$ magnitudes. $V$ was preferred compared with the $B$ values as
their photometric errors are smaller at equal magnitudes.

\begin{figure}
\centerline{
\includegraphics[width=\hsize]{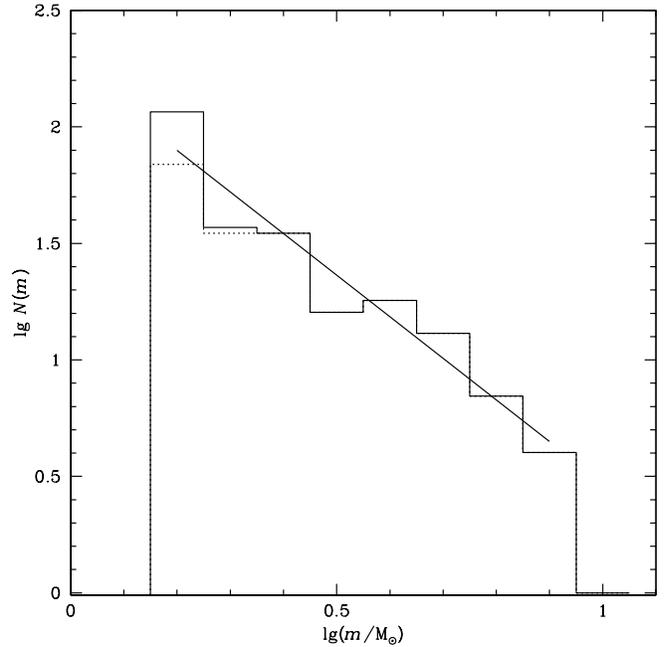}
}
\caption[]{\label{n0581imfplot} Initial mass function of NGC\,581.
  The solid line corresponds to the completeness corrected values, the dotted
  line to the original data. The IMF slope calculated with linear regression
  to the histogram is determined to be $\Gamma=-1.80 \pm 0.19$}
\end{figure}

We included all stars with a mass of higher than $m=1.41 M_\odot$
(corresponding to the previously mentioned value of $V= 17.2 \mbox{ mag}$ or
$\log m=0.15$). In this range, the completeness is $60 \%$ or higher. 198
stars were selected by this criterion. The stars were divided in $\Delta \log
m=0.1$ wide bins. The star numbers were corrected using the results
of the completeness calculation, resulting in 247 ``virtual'' stars from $9.45
M_\odot$ down to $1.41 M_\odot$. The resulting histogram is presented in
Fig. \ref{n0581imfplot}. The single star with a mass around $9.45 M_{\odot}$
($\log m =0.98$) was not taken into account for the IMF determination, since
this single object at one end of the histogram might heavily influence the IMF
slope. The slope of the IMF was determined by linear regression to the
histogram.

We obtained $\Gamma=-1.80 \pm 0.19$ from the stars with a mass from $1.41
M_{\odot}$ to $8.70 M_\odot$ (An experiment including the $9.45 M_{\odot}$
star led to $\Gamma=-2.11 \pm 0.23$ and therefore to a difference in the
exponent of more than 0.3!). The resulting IMF is shown in
Fig. \ref{n0581imfplot}.

\section{Summary and discussion}

NGC\,581 is a young open star cluster with an age of 16 Myr. It is
located at a distance of 2900 pc from the Sun. We derived proper motions of
228 stars in the region of the cluster down to $14 \mbox{ mag}$. 77 of those
can be considered members of the object. A study of the IMF of NGC\,581 leads
to a power law with a slope of $\Gamma=-1.80$.

The results of the photometry mainly coincide with the findings of
Phelps \& Janes (\cite{phelps1}), however, we do not find evidence for a star
formation over a period of as long as 10 Myr: While Phelps \& Janes claim the
necessity of two isochrones of different ages to fit both the blue and red
bright stars in their CMD, the Geneva $\log t=7.2$ isochrone fits all features
of our CMD sufficiently well.

The IMF slope might be slightly higher than Scalo's (\cite{scalo2}) figure for
this mass range, as Phelps \& Janes (\cite{phelps2}) find a steep slope for
NGC\,581, too. Their value of $\Gamma=-1.78$ marks the steepest IMF of their
entire study of open star clusters. Although their IMF is based on a
photometry with a deeper limiting magnitude, our determination has its
advantages, too: Phelps \& Janes only used statistical field star subtraction
and no information of proper motions, which may have great impact on the high
mass range and therefore might affect the slope of the IMF. Furthermore, they
only took one single region in the vicinity of their star clusters for the
field star subtraction, while we were able to use the surrounding of the
cluster itself. And finally, their field of view is so small that they
probably did not cover the entire cluster which, assuming the presence of mass
segregation, might have flattened their IMF due to a lack of low mass stars
from the outer regions of the cluster.

If we would not have based part of the membership determination on the proper
motions, we would have ended up with 9 more stars in the magnitude range
of $V<14 \mbox{ mag}$ (see Fig. \ref{n0581vergleich}). Since the total star
number in this region is low, this would have decreased the IMF the slope to
$\Gamma=1.73 \pm 0.28$. Within the errors, this result would still match with
our value above, however, the shape of this IMF is less well represented by a
power law, which is expressed by the higher error. Therefore we claim that
adding the information of the field star subtraction did improve the
reliability of our IMF study.

\acknowledgements

The authors acknowledge Wilhelm Seggewiss for allocating time at Hoher List
Observatory and Albert Bruch for time at the M\"unster PDS machines. Thanks a
lot to Andrea Dieball for the field star subtraction software and to Klaas
S.\,de Boer and Andrea Dieball for carefully reading the manuscript of this
publication. J.Sa. especially thanks Thomas Schimpke for introducing him to
the use of PDS and the software for handling the digitised photographic plates
and Georg Drenkhahn for his support in programming IRAF
scripts. J.B. acknowledges financial support from the Deutsche
Forschungsgemeinschaft under grant ME 1350/3-2.

\end{document}